\documentstyle[aas2pp4]{article}

\newcommand {\ASCA}{{\em ASCA}}

\slugcomment{To be submitted to ApJ}

\lefthead{H. Murakami et al.}
\righthead{An X-ray Reflection Nebula Sgr B2}

\begin{document}

\title{{\em ASCA} Observations of the Sgr B2 Cloud: \\ An X-Ray Reflection 
Nebula}

\author{Hiroshi~Murakami\altaffilmark{1},
Katsuji~Koyama\altaffilmark{2}, Masaaki~Sakano\altaffilmark{1}, and
Masahiro~Tsujimoto}
\affil{Department of Physics, Faculty of Science, Kyoto University,
Sakyo-ku, Kyoto 606-8502, Japan; hiro@cr.scphys.kyoto-u.ac.jp,
koyama@cr.scphys.kyoto-u.ac.jp, sakano@cr.scphys.kyoto-u.ac.jp, 
tsujimot@cr.scphys.kyoto-u.ac.jp}

\and

\author{Yoshitomo~Maeda\altaffilmark{1}}
\affil{Department of Astronomy and Astrophysics,
The Pennsylvania State University,
University Park, PA 16802-6305, U.S.A.; maeda@astro.psu.edu}
\altaffiltext{1}{Research Fellow of the Japan Society for the Promotion 
of Science (JSPS)}
\altaffiltext{2}{CREST, Japan Science and Technology Corporation
(JST), 4-1-8 Honmachi, Kawaguchi, Saitama, 332-0012, Japan}

\begin{abstract}
 We present the {\ASCA} results of imaging spectroscopy of the giant
molecular cloud Sgr~B2. The X-ray spectrum is found to be very
peculiar; it exhibits a strong emission line at 6.4 keV, a low energy
cutoff below about 4 keV and a pronounced edge-structure at 7.1 keV. The
X-ray image is extended and its peak position is shifted from the core
of the molecular cloud toward the Galactic center by about 1--2
arcminute. The X-ray spectrum and the morphology are well reproduced
by a scenario that X-rays from an external source located in the
Galactic center direction are scattered by the molecular cloud Sgr~B2,
and come into our line of sight. Thus Sgr~B2 may be called an {\it
X-ray reflection nebula}. Possible implications of the Galactic center
activity related to this unique source are presented.
\end{abstract}

\keywords{Galaxy: center --- Galaxy: abundances --- ISM: clouds 
--- ISM: individual (Sgr B2) --- X-rays: ISM}

\section{Introduction}
X-rays from the Galactic center (GC) region have been repeatedly
observed with past X-ray instruments. One of the remarkable
discoveries is a large-scale hot plasma of about 100-pc in diameter,
which is associated with prominent K$\alpha$ lines of He- or H-like
irons (Koyama et al. 1989\markcite{Koyama89}; Yamauchi et
al. 1990\markcite{Yamauchi90}). The total energy of the large-scale
plasma is about $10^{54}$ erg, with the dynamical age of $\sim 10^5$
yr. The {\ASCA} satellite, with imaging capability in the wide energy
band (0.5--10 keV) and reasonable energy resolution, has further found
a high temperature plasma inside the Sgr~A shell (an oval region of
$\sim 2'\times3'$), with the total energy of $3 \times 10^{50}$
erg (Koyama et al. 1996\markcite{Koyama96}).   Using the size of
$\sim 4$ pc and temperature of $\sim 10$ keV (Koyama et
al. 1996\markcite{Koyama96}), we estimate the dynamical age of the smaller 
plasma to be $\sim 10^{3}$ yr.

{\ASCA} also found diffuse emission from the 6.4-keV neutral iron line; 
the brightest region is located over the giant molecular cloud Sgr~B2, but
its X-ray peak is shifted toward the GC from that of the molecular gas
(Koyama et al. 1996\markcite{Koyama96}). Koyama et
al. (1996)\markcite{Koyama96} speculated that the Galactic nucleus Sgr~A$^*$ 
had been bright until some hundreds of years ago, the light travel time between
Sgr~B2 and Sgr~A$^*$, but is currently dim. This putative activity in
the past may also explain the oval-shape plasma surrounding 
Sgr~A$^{*}$, because the dynamical age of the plasma is about 10$^{3}$
yr.

The 6.4-keV line regions, together with the hot plasma surrounding
Sgr~A$^{*}$, may thus provide a challenging scenario for the past
activity of the GC. However the previous report was based on
limited data sets, hence their interpretations were rather preliminary and
qualitative. We therefore have analyzed the X-ray spectrum and
morphology of the Sgr~B2 cloud in further detail combining all the
available data, and try to give more quantitative results and
implications.

With the combined analysis of the CO molecular lines and the
far-infrared dust emissions, the mass of Sgr~B2 is estimated to be 6
$\times 10^6$ M$_\odot$ within a region of $\sim$ 45 pc in diameter
(Lis \& Goldsmith 1989\markcite{Lis89}), hence it is one of the
largest molecular clouds in the Galaxy.

We assume the distance to Sgr~B2 is the same as that to the GC (Sgr~A$^*$)
or 8.5 kpc, which is within an error of estimated distance 
of 7.1 $\pm$ 1.5 kpc (Reid et al. 1988\markcite{Reid88}).
Then the  distance between Sgr~B2 and Sgr~A$^*$ is about 100 pc.

\section{Observations}
Two observations of Sgr~B2 were made with {\ASCA} on 1993 October 1,
and on 1994 September 22--24. In both observations, all four
detectors, two Solid-state Imaging Spectrometers (SIS0, SIS1) and two
Gas Imaging Spectrometers (GIS2, GIS3) were operated in parallel,
hence four independent data sets were provided. Details of the
instruments, the telescopes and the detectors, are found in 
Tanaka, Inoue, \& Holt (1994)\markcite{Tanaka94}, Serlemitsos et
al. (1995)\markcite{Serlemitsos95}, Burke et
al. (1991)\markcite{Burke91}, Ohashi et
al. (1996)\markcite{Ohashi96}, Makishima et
al. (1996)\markcite{Makishima96}, and Gotthelf
(1996)\markcite{Gotthelf96}. Each of the GIS was operated in PH
mode with the standard bit-assignment. The SIS data were obtained in
4-CCD bright mode. The data were post-processed to correct for
spatial gain non-linearity. Data taken at geomagnetic cutoff
rigidities lower than 6 GeV c$^{-1}$, at elevation angles less than
10$^\circ$ for the GIS and 5$^\circ$ for the SIS from the Earth rim, or during
the passage through the South Atlantic Anomaly were excluded. For the SIS,
we also excluded the data taken at elevation angles from the bright
Earth rim less than 25$^\circ$. After these filters were applied, the
net observing times were  95 ksec for the GIS and 85 ksec for the SIS.

\section{Results}
Koyama et al. (1996)\markcite{Koyama96} have already reported that the
Sgr~B2 cloud region is particularly bright in the 6.4-keV line. We
therefore made X-ray images in narrow energy bands with a central
energy of 6.4 keV and width of twice the energy resolution (FWHM):
5.8--7.0 keV for the GIS and 6.2--6.6 keV for the SIS. Figure 1 shows the
narrow band SIS (a) image laid over the radio intensity contours of
the CH$_3$CN line (Bally et al. 1988\markcite{Bally88}) and the GIS
(b) image. Since the SIS image is already found in Figure 3b in Koyama et
al. (1996)\markcite{Koyama96}, we present the combined image of the two
observations.

\subsection{Spectrum of Sgr B2}
For the X-ray spectrum, we mainly used the GIS data, because,
in the high energy band including the iron K-shell line, the GIS
provides better statistics than the SIS. 
The GIS spectrum given in Figure 2 is obtained by summing the X-ray
photons in $3'$-radius circles around the X-ray peaks of the GIS
images. For the background spectrum, we used an
elliptical region with the major axis parallel to the Galactic plane,
excluding the region of Sgr~B2 (a $3'$-radius circle) and the other
X-ray bright spot (the other $3'$-radius circle) at the west of
Sgr~B2. The source and the background regions are shown in Figure 1b
with the solid circle and dotted ellipse, respectively.

In order to derive quantitative feature of the Sgr B2 X-rays, we fit
the spectrum to two phenomenological models, both a thermal
bremsstrahlung and a power-law model, each with a Gaussian line.  We
used the Morrison \& McCammon (1983)\markcite{MorMc83} cross section
for the absorption. Due to large absorption at low energy, the
available data to be fitted are in the 4.0--10.0 keV band. However the
limited energy band and rather poor statistics do not allow us to
constrain either model. Therefore we assumed a power-law of (fixed)
photon index 2.0 (Koyama et al. 1996\markcite{Koyama96}). The best-fit
parameters are given in Table 1. The 6.4-keV line, as we expected, is
very strong with an equivalent width of 2.9 keV. The hydrogen column
density is $N_{\rm H} \sim 8 \times 10^{23}$ H cm$^{-2}$, and the
luminosity is $\sim 10^{35}$ erg s$^{-1}$ (Here and elsewhere, all
X-ray luminosities are corrected for absorption unless otherwise
noted).

We also made a SIS spectrum and fit it with the same model of the
GIS. However, since the limited photon statistics prevented us from
determining the absorption depth, we fixed the
hydrogen column to the best fit value for the
GIS. The results are given in Table 1. The SIS results were found to
be consistent with the GIS. The narrow field of view of SIS severely
constrains the selection of the background region for the SIS spectrum
and increases background-subtraction uncertainties. Here and after, we
thus analyzed only the GIS spectrum for more detailed studies.

Koyama et al. (1996)\markcite{Koyama96} reported that the most
prominent iron K-shell lines from the GC region are the 6.4-keV line
from cold iron and the 6.7-keV line from He-like iron. 
Accordingly, we fit the spectrum to the power-law and two narrow Gaussian 
lines at 6.4 and 6.7 keV, and found no significant line at 6.7 keV, with the 
flux less than
10\% of the 6.4-keV line.  Thus the observed line profile is reproduced by the 
6.4-keV line alone. 

The observed hydrogen column of $8 \times 10^{23}$ H cm$^{-2}$ is at
least 5 times larger than that of interstellar gas to the GC region
(Sakano et al. 1997\markcite{Sakano97}). This means that the large
absorption column is due to local gas near or at the Sgr~B2 cloud.

We found a deep iron edge in the spectrum. We fit the spectrum
allowing the iron column density to be free.
Then the iron column density is estimated to be 4$\times 10^{19}$ Fe
cm$^{-2}$.

\subsection{X-Ray Morphology}
To determine accurate X-ray peak positions, we made projected
X-ray intensity profiles (6.2-6.6 keV) on the right ascension (RA) and
declination (Dec) axes, and fit to a phenomenological model
function: Gaussian plus exponential tails for the two sides. Each of
the best-fit positions of the X-ray peak thus determined in the GIS
and the SIS images are RA(2000) $=$ $17^{\rm h}\ 47^{\rm m}\ 20\fs5$,
Dec(2000) $=$ $-28\arcdeg\ 24'\ 20''$ and RA(2000) $=$ $17^{\rm h}\
47^{\rm m}\ 20\fs3$, Dec(2000) $=$ $-28\arcdeg\ 24'\ 13''$,
respectively, with statistical errors of $10''$ for the GIS and $15''$
for the SIS. Although the SIS has a better spatial resolution than the
GIS, it shows a larger error, due mainly to the poor photon statistics
compared with the GIS.  An additional but even larger error is
uncertainty in the {\ASCA} attitude determination  which is less than
$40''$ (Gotthelf 1996\markcite{Gotthelf96}).

The radio peak of Sgr~B2 is RA(2000) $=$ $17^{\rm h}\ 47^{\rm m}\
20\fs1$, Dec(2000) $=$ $-28\arcdeg\ 23'\ 6''$ in $4\farcs5 \times
3\farcs7$ beam size from the HC$_3$N observation by Lis et
al. (1993)\markcite{Lis93}. Thus the X-ray peak is shifted from the
cloud core to the south by $1'.2$, which is significantly larger than
the X-ray position errors.

Spatial extent of the 6.4-keV line and the continuum X-ray images are
examined with the SIS data, because spatial resolution of the SIS is
better than that of the GIS. We made radial profiles for the 6.2--6.6
keV and the 4.0--10.0 keV (excluding the 6.2--6.6 keV band) bands; the
former band is dominated by the 6.4-keV line photons, while the latter
represents the continuum X-ray photons. For comparison we 
simulated the point spread function (PSF) by the ray-tracing code 
(Kunieda et al. 1995\markcite{Kunieda95}). Then we fit the PSF
to the observed radial profiles with one free parameter of normalization, 
and found the best-fit reduced $\chi^2$ to be 1.75 (40 d.o.f.) for the 6.4-keV
line and 3.46 (40 d.o.f.) for the continuum band. These reduced $\chi^2$
values exclude the possibility of Sgr B2 being a point source.
The observed radial profile and the PSF are shown in Figure 3.

It is conceivable, though, that significant fractions of X-rays from a
point source near or behind the GC region are scattered by dust grains
and produce an X-ray halo, which mimics this source to be an extended
object. Within the {\ASCA} spatial resolution, however, no point
sources near the GC region are found to be extended, as is
demonstrated in Maeda et al. (1996\markcite{Maeda96}) and Sakano et
al. (1999\markcite{Sakano99}). Thus scattering by interstellar dust is
negligible.

Since the column density to Sgr~B2, $N_{\rm H} = 8 \times 10^{23}$ H
cm$^{-2}$ is at least 5 times larger than that of interstellar gas to
the GC region (Sakano et al. 1997\markcite{Sakano97}), the X-ray halo
due to dust grains in the Sgr~B2 cloud would be more enhanced. With a
distance $D$ to a point source, a halo size by dust-scattering at the
Sgr~B2 cloud is estimated to be $2\farcm7~\times$~(1$-$8.5kpc/$D$)
(Xu, McCray, \& Kelly 1986\markcite{Xu86}; Mitsuda et
al. 1990\markcite{Mitsuda90}), while the scattered X-rays in the halo
are estimated to be about 25 \% of the direct beam (Mathis \& Lee
1991\markcite{Mathis91}; Mauche \& Gorenstein
1986\markcite{Mauche86}). Taking an extreme case, where a point source
is at the infinite distance to make the scattered halo the most
extended, we simulated the {\ASCA} profile of a point source plus X-ray
scattered halo, and fit it to the observed profile. This hypothesis
still leads to smaller extension than observed, and is excluded with
the best-fit reduced $\chi^2$ of 1.45 (40 d.o.f) and 2.84 (40 d.o.f)
for the 6.4 keV and continuum bands, respectively.

We thus conclude that the extended structure we
observed is not due to an X-ray halo of a point source, but is
intrinsic feature of the Sgr~B2 cloud.

\section{X-Ray Reflection Nebula }
In section 3, (1) we confirmed the presence of the very strong 
emission line at 6.4 keV, (2) we found a large low-energy cutoff and a 
deep absorption edge at 7.1 keV, both requring an extremely large column 
near or at the Sgr~B2 cloud, and (3) we found that the X-ray emitting region 
is extended with its peak position about 1--2 arcminute offset from the cloud 
center to the GC side.

This peculiar X-ray spectrum and morphology 
are attributable to Thomson scattering (continuum
emissions), photo-electric absorption of neutral iron atoms
(low-energy cutoff and iron K-edge), and fluorescence (6.4-keV line),
produced by an irradiation of an external X-ray source. 
We refer this scenario as ``X-ray Reflection Nebula''(XRN) model.
This section is devoted to numerical simulations
to see   whether or not the XRN model 
reproduces the X-ray spectrum and morphology of Sgr~B2.
 
\subsection{Numerical Simulations}
The geometry and configuration for XRN simulations are
given in Figure 4, with the following assumptions.

(1) A model cloud (an XRN) is cylindrically symmetric
with its axis parallel to our line of sight.
The mass distribution of the cloud is taken from 
the result of $^{13}$CO and the 
C$^{18}$O observations by Lis \& Goldsmith (1989)\markcite{Lis89};
\begin{equation}
\left( \frac{n_{{\rm H}_2}}{1 {\rm cm}^{-3}} \right)
=5.5 \times 10^4 \left( \frac{r}{1.25{\rm pc}} \right)^{-2} + 2.2 \times
10^3,
\end{equation}
where $n_{{\rm H}_2}$ is the number density of hydrogen molecules, and $r$ is 
the
projected distance from the center of the cloud. This formula gives
the total molecular mass to be $6.3 \times 10^6$~M$_\odot$ in the
45-pc diameter region.

(2) A primary source (irradiating source) is located in the GC-side
with the normal angle to the cloud (cylinder) axis. The spectrum of
the primary source is a power-law with photon index of 2 (Koyama et
al. 1996\markcite{Koyama96}).

For  numerical simulations, we divided the model cloud into 
$225\times225\times113$
cells (each cell is 0.2 pc in size), and calculated absorbed, reflected, and
fluorescent X-rays in each cell as a function of incident X-ray
energy $(E)$ from the primary source.  For simplicity we adopted further
assumptions.

(3) The reflection and fluorescence are isotropic, and the absorption
is given by the analytical function of the photoelectric cross section 
for the standard interstellar 
matter (Morrison \& McCammon, 1983\markcite{MorMc83}). 
The cross section of an iron atom was taken 
from Henke et al. (1982) \markcite{Henke82}. 

(4) Since the Thomson scattering optical depth is much smaller than
that of the photo-electric absorption in the relevant energy band below 10 keV,
we neglect the multiple Thomson scattering. 

Then the probability $P_{\rm scat} (E)$ that a primary X-ray photon arrives at
a cell, is scattered, and comes into our line of sight, is given as a
function of photoelectric absorption cross section ($\sigma(E)$),
the column density from the primary source to the cell ($N_{\rm
H}^a$), the sell size ($l$), Thomson scattering cross section
($\sigma_{\rm T}$), the Hydrogen number density in the cell
($n_{\rm cell}$), and the column density from the cell to the surface of
the cloud ($N_{\rm H}^b$) (see figure 4).

\begin{equation}
\label{scat}
P_{\rm scat} = l n_{\rm cell} \sigma_{\rm T} 
\exp{\it (- N_{\rm H}^a \sigma(E) -N_{\rm H}^b \sigma(E))}.
\end{equation}

Some fraction of the primary X-rays with energies above 7.1 keV, the
K-edge of neutral iron, are absorbed with the cross section of an iron
atom ($\sigma_{\rm Fe}(E)$).  The absorbed X-rays are re-emitted as
the 6.4-keV photons, with the probability of the fluorescence yield of
0.34 (Bambynek et al. 1972\markcite{Bamby72}).

Thus the probability $P_{6.4}$, that a primary X-ray ($ E \geq $ 7.1 keV)
is converted to the 6.4 keV line and comes into our line of sight, is similarly
given;

\begin{equation}
\label{6.4}
P_{6.4} = 0.34 l n_{\rm cell} Z_{\rm Fe}
\sigma_{\rm Fe}(E) \exp{\it (- N_{\rm H}^a \sigma(E) -N_{\rm H}^b
\sigma({\rm 6.4}))},
\end{equation}
where $Z_{\rm Fe}$ is abundance of iron.

In both the cases, the interstellar absorption from the primary source to
the cloud is neglected. Adopting the Galactic absorption from the
cloud to the observer of 1 $\times 10^{23} {\rm H} \ {\rm cm}^{-2}$
(Sakano et al. 1997\markcite{Sakano97}), we finally obtained simulated
spectra and images in the continuum and the 6.4-keV line bands.
 
\subsection{Spectrum}
The simulated spectra are convolved with the response function and are
compared with the observed spectrum. We first fit the observed spectrum
to the simulated one with the XRN model of solar abundances, where a free
parameter is only the normalization of the flux.

This simulated spectrum, however, is rejected with a reduced $\chi^2$ of 
2.35 (36 d.o.f.). 
Large residuals are found in the flux of the 6.4 keV line, depth of
the K-edge and low-energy absorption. We accordingly vary
the abundances collectively, fixing the relative ratio to be solar. 
We find an acceptable fit with the reduced-$\chi^2$ of 1.18 (35
d.o.f.), when the abundances are 2.2 solar. 

Since we have already found the deep edge of iron at 7.1 keV, we further
search for a better XRN model allowing the iron abundance to be an
additional free parameter, and find a better fit with the
reduced-$\chi^2$ of 0.78 (34 d.o.f.).
The best-fit XRN spectrum is given in Figure 2 by the solid line.
The abundances of iron and the others are determined respectively to
be 2.4 and 1.6, while the 90\% error contour is given with the
solid-line ellipse in Figure 5a (see section 4.3, 4.4).

\subsection{X-ray images}
With the same procedure for the XRN spectral model, we simulate
XRN images of the 6.4-keV fluorescent line (Figure 6) and the
continuum. The XRN images are convolved with the point spread function
produced by the ray-tracing code (Kunieda et
al. 1995\markcite{Kunieda95}). Then we make radial profiles of the
XRN model, and fit to the observed profiles given in Figure 3.
Allowing normalization and background level as two free parameters, we
get acceptable fits with the reduced-$\chi^2$ values of 1.25 (40
d.o.f.) for the 6.4 keV line and 1.06 (40 d.o.f.) for the continuum
X-rays. The best-fit curves are shown in Figure 3 with the dotted
lines.

Since the radial profile is less sensitive to the abundances and we
already obtained acceptable fits with the solar abundances, no
significant constraint on the abundances is given. We hence try to
give possible constraint on the abundances using the separation angle
between the 6.4-keV X-ray peak and the cloud core, because larger
absorption in the molecular cloud gives further shift of the 6.4 keV
peak to the primary source side.

The observed 6.4 keV peak is $3 \pm 2$ pc away from the radio peak. We
thus make constant-separation lines of 1 pc and 5 pc by XRN models,
allowing abundances of iron and the other elements to be free
parameters. The results are given in Figure 5a (see section 4.4) with
solid lines. We find, however, no overlapping region in the abundances
of iron and the other elements between those estimated with the XRN
spectrum and images (separation angle).

\subsection{Mass and Abundance}
The disagreement of the best-fit abundances between the XRN images
and spectrum may be attributable to an improper assumption on the gas
distribution for the Sgr~B2 cloud, either in the shape, total gas
mass, or probably both. However, for simplicity, we make XRN spectra and images
and fit the data allowing the total gas mass and abundances to be
free parameters.

Consistent results are obtained by reducing the total gas mass to be
$<$ 1/2 of the initial gas assumption of Lis \& Goldsmith
(1989)\markcite{Lis89}. For example, we show the case of total gas
mass of 1, 3/4, 1/2, 1/4 of Lis \& Goldsmith (1989)\markcite{Lis89} in
Figure 5. The allowable regions are given by solid lines and ellipse
of each graph. We constrain the total mass and the abundance of this
cloud to be $<$ $3 \times 10^{6}$ M$_\odot$ and $>$ 2 solar
respectively.

Thus the {\it X-ray reflection nebula} scenario has a potentiality
to determine the total gas mass and chemical
abundances. However, since the observed value of the peak separation
still has a large error, further study along this approach should be
postponed until high quality data in morphology and spectroscopy 
become available.

\subsection{Irradiating Source}
Using the best-fit parameters given in Figure 5, and from the
equations 2 and 3, we calculate the fractions of the reflected and
fluorescent fluxes, hence estimate the required luminosity of the
primary source to give the observed fluxes of continuum X-rays and the
6.4-keV photons. The required luminosity of the primary X-ray source
is,
\begin{equation}
L_{\rm 2-10 keV} \sim 3 \times 10^{39} \left( \frac{d}{100 {\rm
pc}} \right)^2 {\rm erg\ s}^{-1},
\end{equation}
where $d$ is the distances from the primary X-ray source to
Sgr~B2. Since the reflecting region is
extended largely over the Sgr~B2 cloud, the required X-ray flux
should be the value averaged over a time of about 100 yr. This
may exclude galactic binary transient sources as the primary X-ray, whose duty
cycle of flaring may be very short.  We hence compare the required
flux with the observed luminosities of rather stable sources found in
the catalogue near the GC in Table 2 (Sidoli et
al. 1999\markcite{Sidoli99}).

The observed luminosities of cataloged bright X-ray sources near the
GC are 1--4 $\times 10^{36}$ erg s$^{-1}$, which are much lower than
that required to explain the reflected luminosity of Sgr~B2.

\section{Discussion}
Our interpretation of Sgr~B2 as an XRN is supported by the numerical
simulation stated in section 4. The model well explains the spectrum
and morphology of Sgr~B2, and we can constrain the mass and the
abundances of this cloud to be $<$ $3 \times 10^{6}$ M$_\odot$ and $>$
2 solar, respectively. Since Sgr~B2 is located near the GC, the
chemical abundances would be similar to those of the GC region. It has
been debatable whether the GC region is overabundant or not (Binette
et al. 1982\markcite{Binette82}; Luck 1982\markcite{Luck82}; Shaver et
al. 1983\markcite{Shaver83}; Ratag et al. 1992\markcite{Ratag92};
Sellgren, Carr, \& Balachandran 1997\markcite{Sellgren97}; Ramirez et
al. 1998\markcite{Ramirez98}). Our result favors the overabundance for
the GC region.

The total gas mass of the Sgr~B2 has been poorly known. Oka et
al. (1998) presented two possibilities about the mass in the 42 pc
diameter region to be either $7.1 \times 10^{6}$ M$_\odot$ or $2.0
\times 10^{5}$ M$_\odot$, depending whether the cloud is in
self-gravitational equilibrium or is in pressure equilibrium with the
hot gas and/or the magnetic field near the cloud region.  The former
estimation is supported by Lis \& Calstrom (1994) for instance, who
claimed the mass to be $2 \times 10^{6}$ M$_\odot$ based on the
observation of 800 $\mu$m dust emission.  Our result, on the other
hand, favors the latter estimation.  This apparent inconsistency is
conceivable if the central dense core of the Sgr~B2 is in the
self-gravitational equilibrium and the outer part is in the pressure
equilibrium.

We find  no X-ray source which exhibits enough X-rays to produce the
flux of possible XRN, Sgr~B2. Koyama et al. (1996)\markcite{Koyama96}
pointed out that the hot plasma prevailing the GC region can explain
only 10 \% of the Sgr~B2 flux. Thus, one plausible possibility is, as
already proposed by Koyama et al. (1996)\markcite{Koyama96}, that the
Galactic nucleus Sgr~A$^*$ was bright some hundreds of years ago, the
light travel time between Sgr~A$^*$ and Sgr~B2, and is dim at present.

Sunyaev \& Churazov (1998)\markcite{Sunyaev98} simulated the
morphology of the surface brightness distribution, the equivalent
width and the shape of the fluorescent line from the Sgr~B2 cloud as a
function of time and relative position of the XRN and the primary
(irradiating) source. They examined both the cases of a flare-like and
a step-function (X-ray flux is constant but suddenly decreases to
zero) irradiation by an X-ray source. As they demonstrated, a short
time variability is the key to investigating the nature of the
irradiating source, whether it is the Galactic nucleus or a binary
X-ray source.  For comparison with data, observations with higher
angular resolution are necessary. The {\em Chandra} resolution of 0.5
arcsec, for example, can attain the time resolution of about 1 month,
which is typical flare duration of a transient binary. Future
observations may thus reveal more realistic 2-dimensional gas
distribution.

 If Sgr~A$^*$ was a bright X-ray source in the past, we can expect
other molecular clouds than Sgr~B2 also as the 6.4-keV sources. The
X-ray bright spot near the radio arc is such a candidate. Sgr~C 
at the similar distance of Sgr~B2 from Sgr~A$^*$ 
(on the other side of the center), however, is not distinctly bright 
in the 6.4-keV line (Koyama et al. 1996\markcite{Koyama96}). 
The mass of the Sgr~C cloud is about 1/7
of Sgr~B2 (Oka et al. 1998\markcite{Oka98}), hence the luminosity is
also 1/7.  Thus whether the Sgr~C is an XRN is unclear with
the present photon statistics.
High quality X-ray and radio data of the cloud are crucial to judge
the XRN scenario, to search for past activities of the GC region and
Sgr~A$^*$, or to investigate the mass distribution and abundances of
the molecular cloud.

\section{Summary}
The {\ASCA} observations highlight peculiar characteristics of the giant
molecular cloud Sgr~B2, and our analysis supports an interpretation that
the cloud is an XRN. The observational facts and their implications are 
the following.

\begin{enumerate}
\item We find a strong emission line at 6.4 keV from cold iron with
an equivalent width of about 2.9 keV. No significant emission line
from He-like iron, which is commonly observed near the GC region, is
found from Sgr~B2.

\item The spectrum exhibits a large absorption at energies lower than 4
keV, equivalent to the hydrogen column of $8 \times 10^{23}$ H
cm$^{-2}$ (with the solar abundances) and a pronounced edge-structure
at 7.1 keV with equivalent iron column $N_{\rm Fe}$ of 4 $\times
10^{19}$ Fe cm$^{-2}$.

\item The X-ray image of the 6.4-keV line as well as the continuum
band is found to be extended with the peak shifted toward the GC by
about 1--2 arcminute.

\item The spectrum and the X-ray morphology are well reproduced by an
XRN model, in which a strong X-ray source is irradiating the molecular
cloud Sgr~B2 from the GC direction.

\item The XRN  model favors an overabundance and 
smaller total gas mass of the Sgr~B2 cloud.

\item No X-ray source luminous enough to produce the XRN is found.
One possibility is that the past activity of the GC is responsible for the
presence of reflected X-rays and absence of the irradiating source at present.
\end{enumerate}

\vspace*{1em}

\acknowledgements
The authors express their thanks to all the members of the {\ASCA}
team. H.M, M.S. and Y.M. are financially supported by the Japan Society for
the Promotion of Science. The authors also thank Dr. T. Oka for his
useful discussion. Thanks are also due to Prof. G. Garmire for a
critical reading of the manuscript. The authors are grateful to the
anonymous referee for the insightful comments.

\newpage
\onecolumn

\begin{figure}
\vspace{1.5cm}
\plottwo{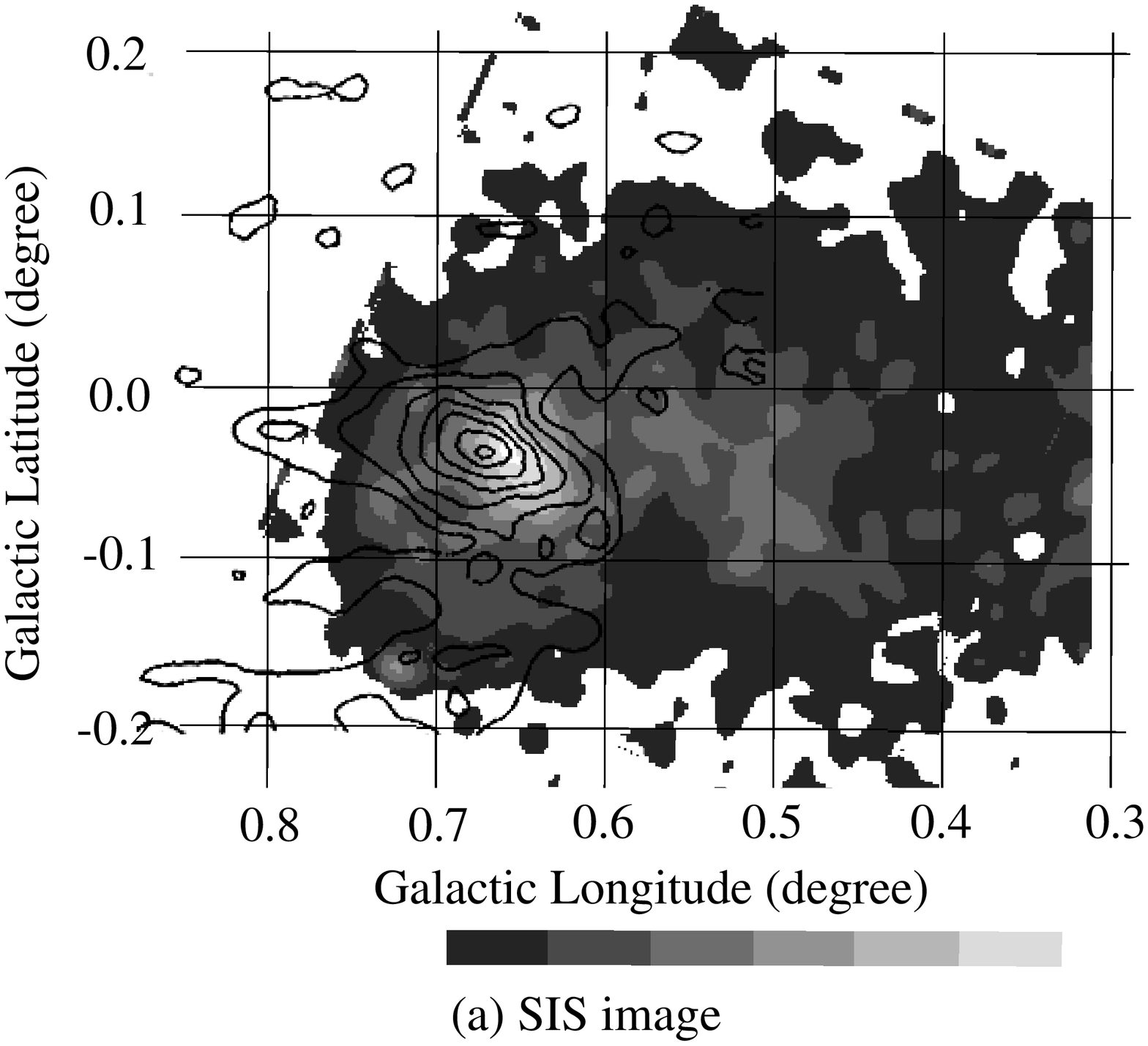}{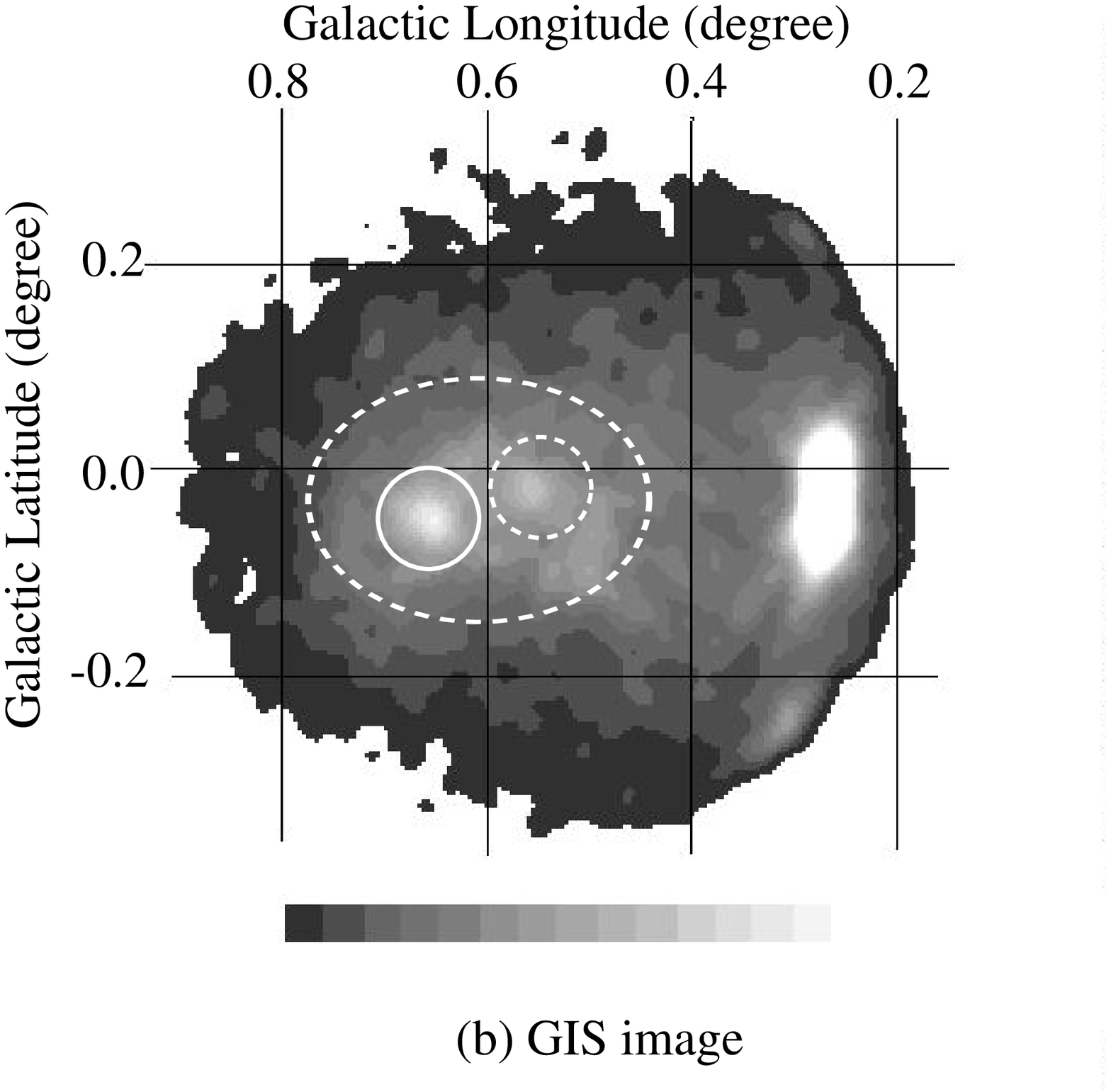}
\caption[f1a.ps, f1b.ps]{(a) The 6.4-keV line image around the Sgr~B2
cloud obtained with the SIS, laid over the CH$_3$CN line contours (Bally
et al. 1988). The 6.4-keV brightness distribution is shifted from the
radio distribution by $\sim 1'.2$ to the Galactic center side (to the
right in the figure). (b) The 6.4-keV line image with the GIS. The source
and the background regions are shown by the solid circle and the
dotted ellipse, respectively. The dotted circle encloses the other
X-ray bright spot, which is excluded from the background region.
The bright source at the boundary is a galactic binary X-ray source
(1E 1743.1$-$2843).}
\end{figure}

\begin{figure}
\epsscale{0.8}
\plotone{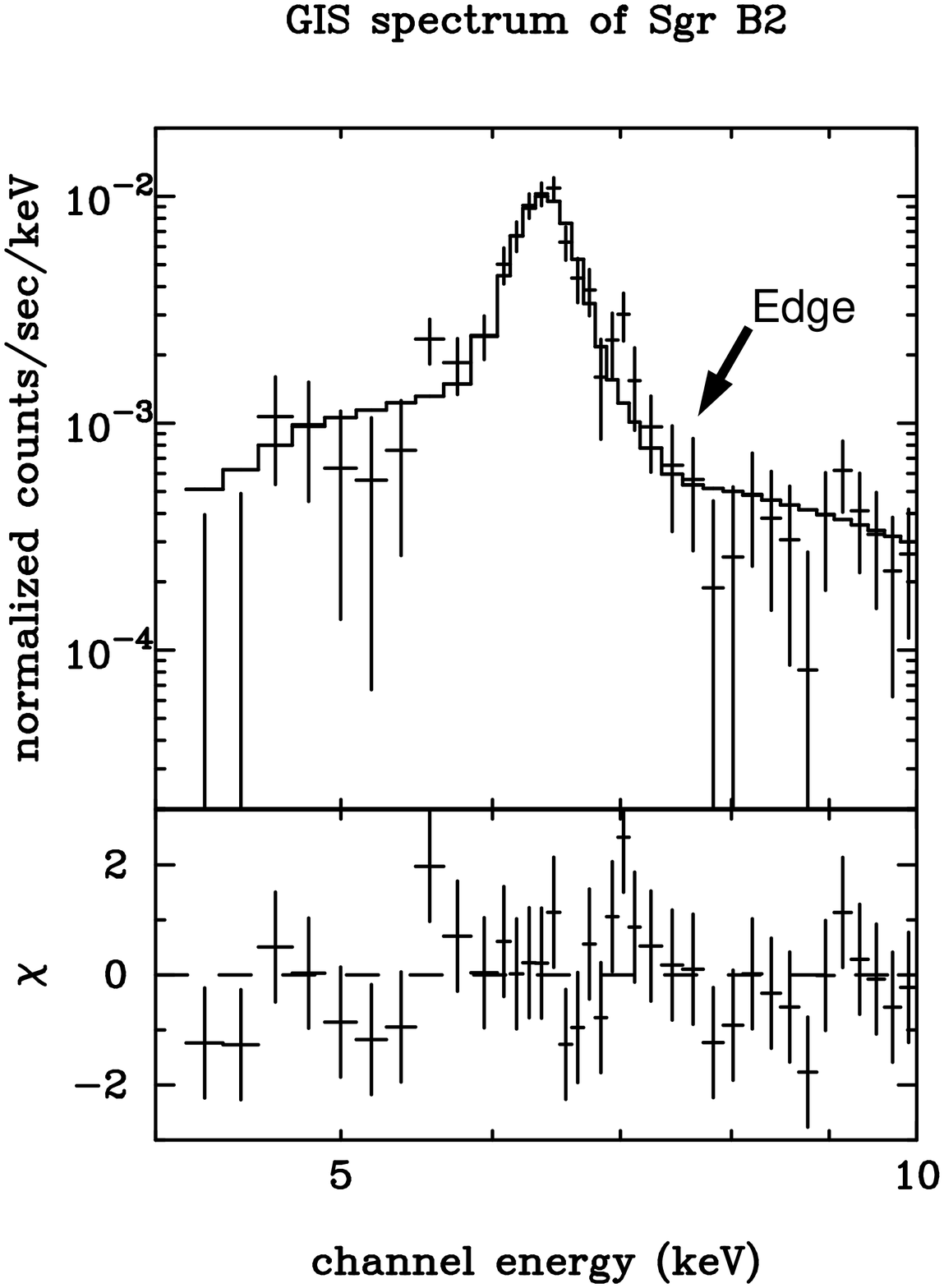}
\caption[f2.ps]{ The GIS (GIS2 + 3) spectrum of Sgr~B2. The solid
line shows the simulated spectrum of an XRN (see section 4.2),
after the convolution of the response function.}
\end{figure}

\begin{figure}
\epsscale{1.0}
\plottwo{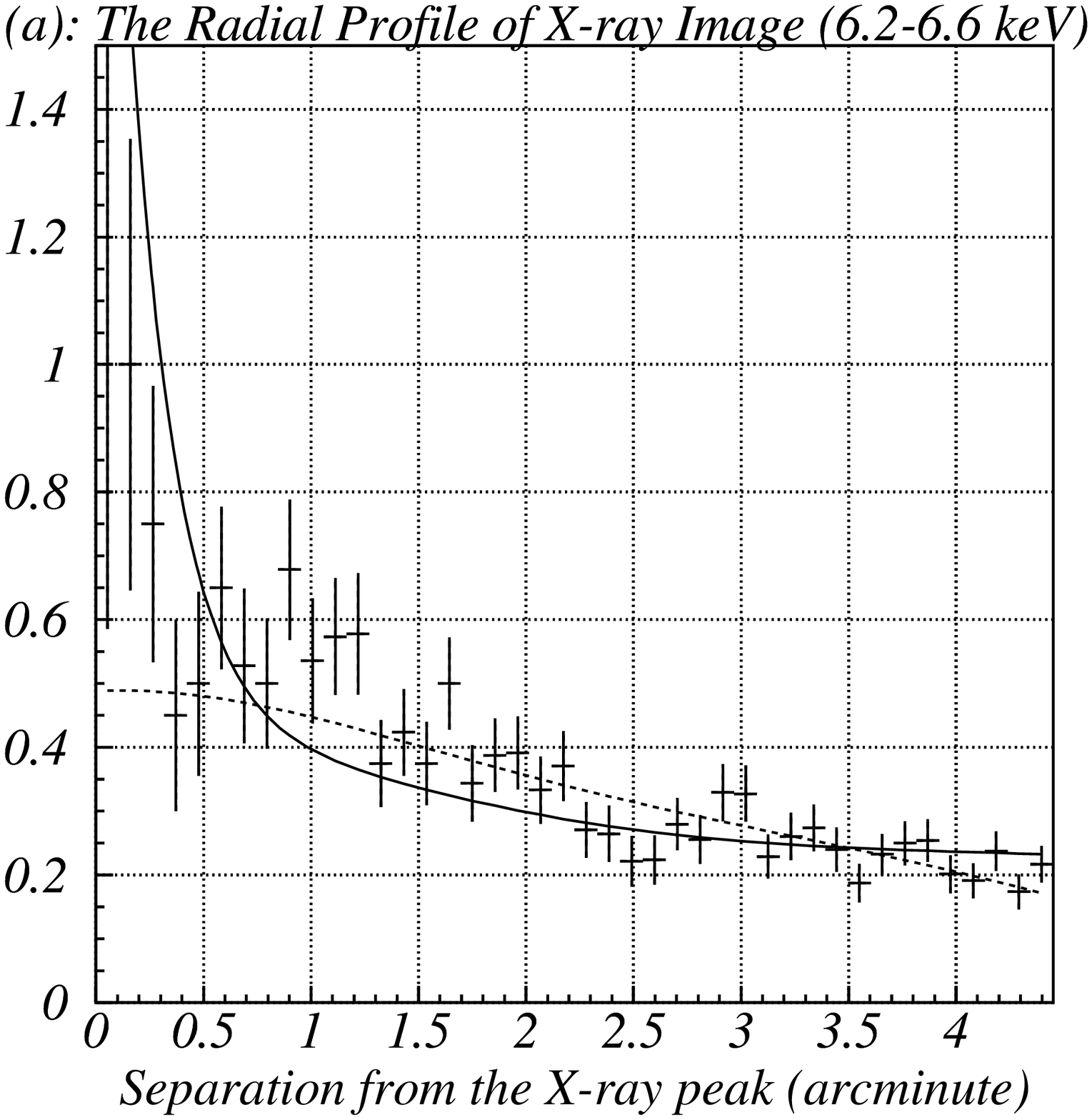}{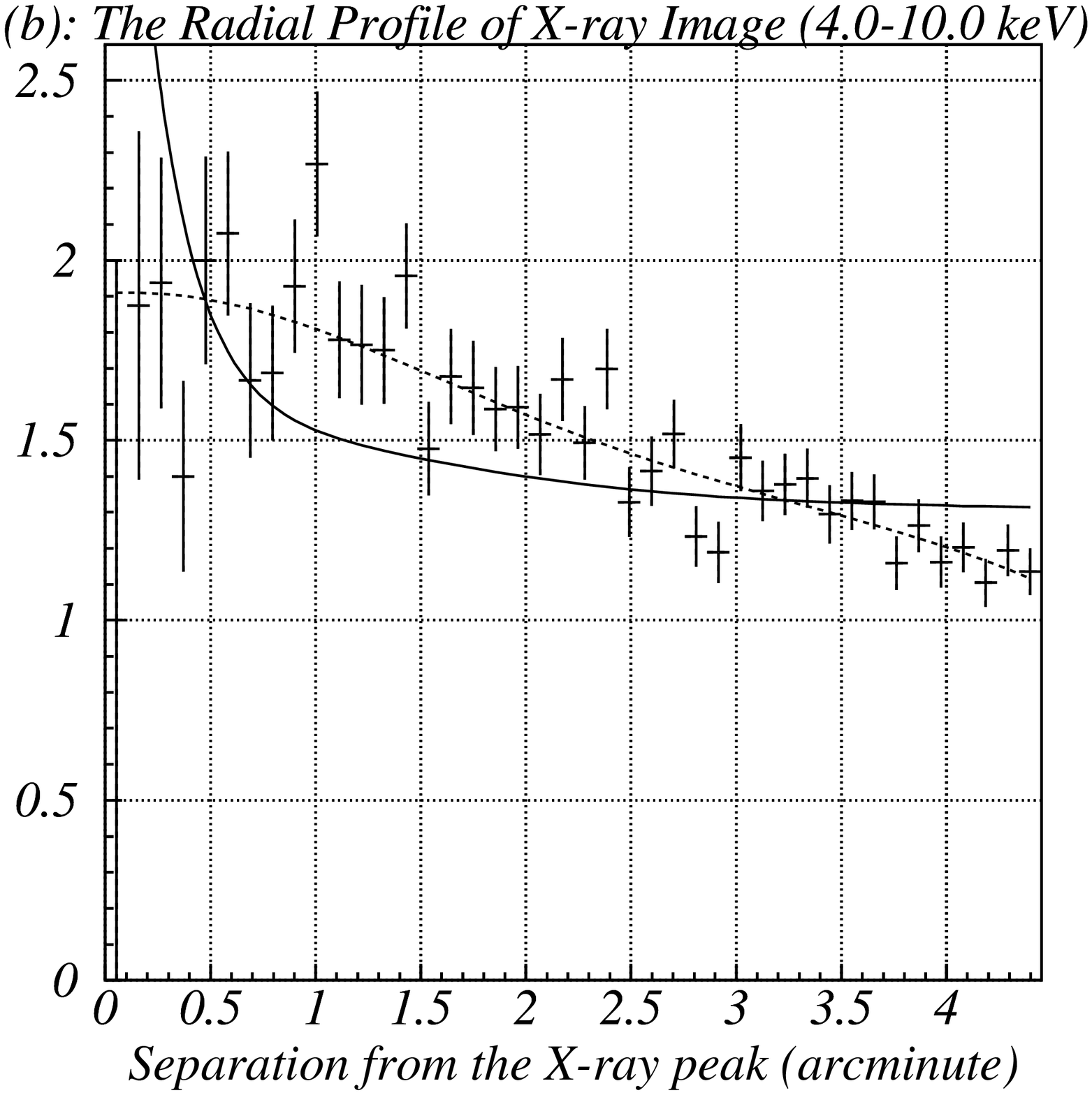}
\caption[f3a.ps, f3b.ps]{The SIS radial profile of X-ray intensity in the
energy range of (a) 6.2--6.6 keV and (b) 4.0--10.0 keV (excluding the
6.2--6.6 keV band). The radial profile of the PSF (point source profile)
is  shown by the solid line, while the simulated XRN profile (see section 4.3)
after the convolution of PSF is given by dotted lines.
}
\end{figure}

\newpage

\begin{figure}[p]
\epsscale{1.0}
\plotone{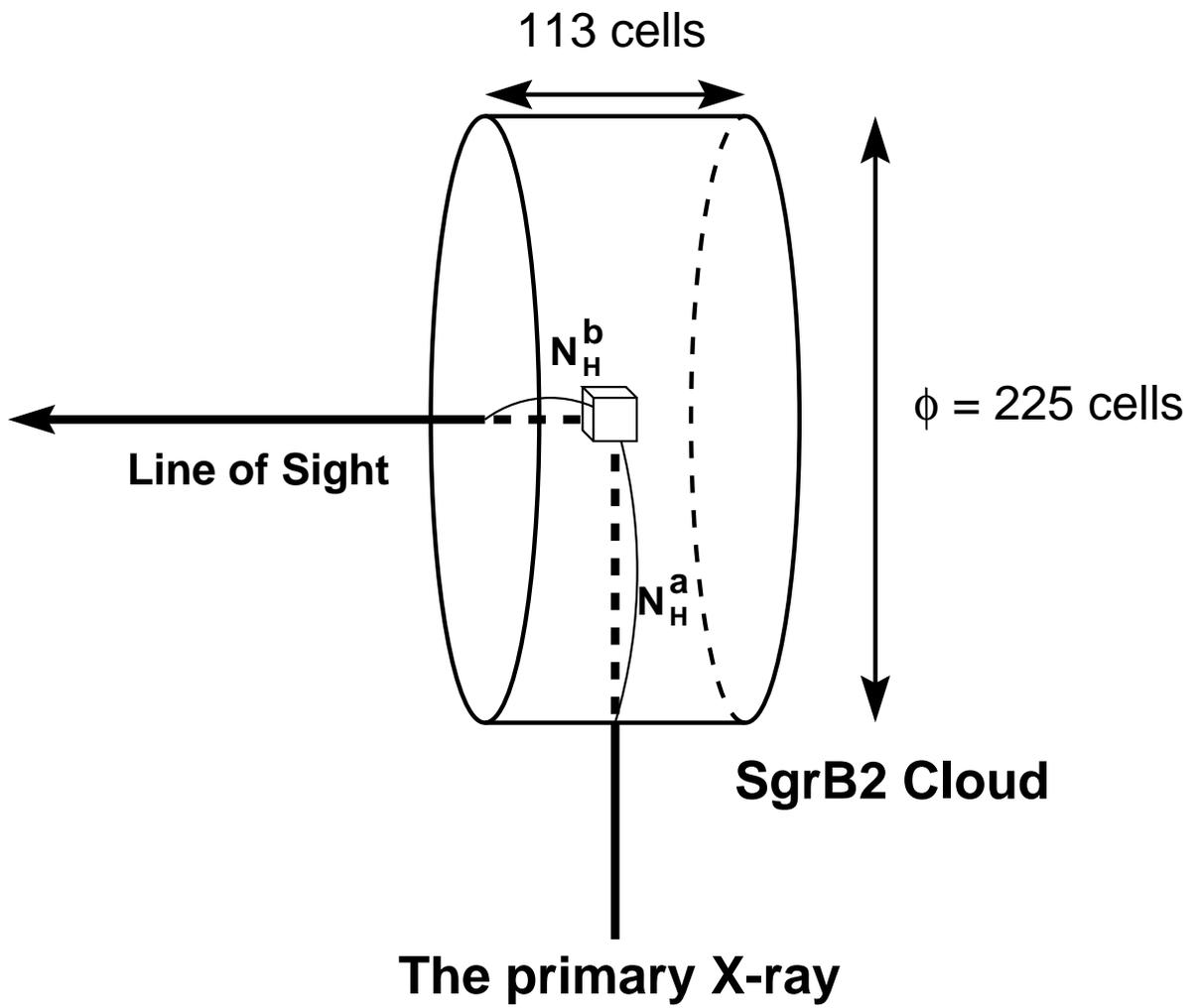}
\caption[f4.ps]{The schematic view of the XRN
simulation (see section 4).}
\end{figure}

\begin{figure}[p]
\epsscale{1.0}
\plotone{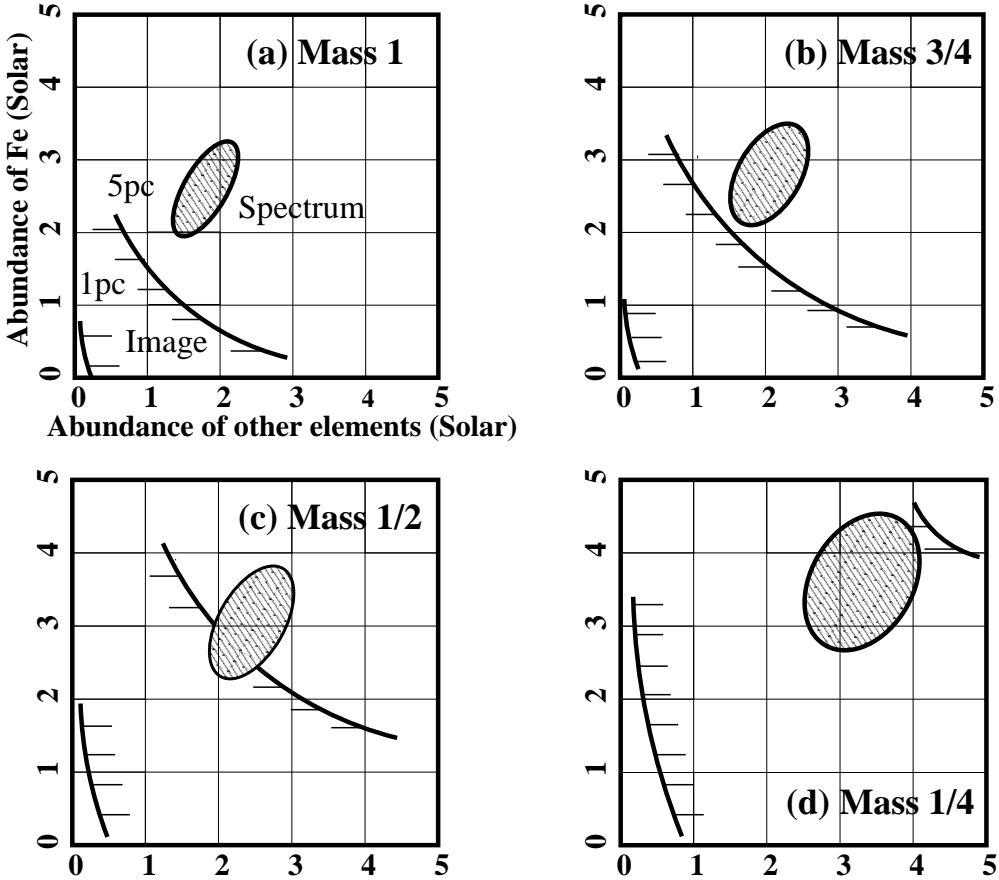}
\caption[f5.ps]{The 90\% confidence contours for the
abundances of iron and other elements obtained by the spectrum analysis
under the four assumptions of the total mass (a: 6.3 $\times 10^6$,
b: 4.7 $\times 10^6$, c: 3.2 $\times 10^6$, d: 1.6 $\times 10^6$ in
unit of solar mass; each corresponds to 1, 3/4, 1/2, 1/4 of the
estimation of Lis \& Goldsmith 1989). 
The constant-separation lines for the 1-pc
and 5-pc between the 6.4-keV X-ray peak and the molecular
core are also illustrated in each case.}
\end{figure}

\begin{figure}[p]
\epsscale{0.9}
\plotone{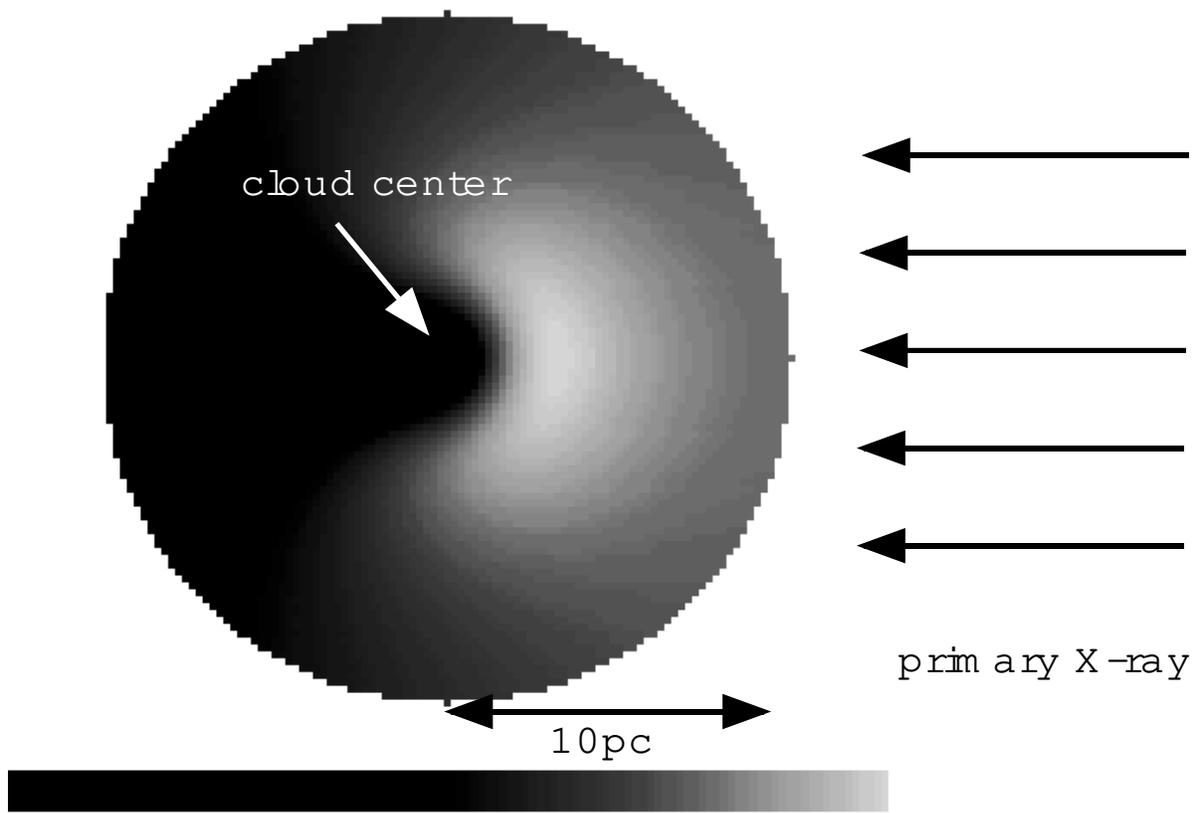}
\caption[f6.ps]{The simulated XRN image of the 6.4-keV line. 
The brightest region is shifted from the cloud center to the 
side of the primary X-ray source.}
\end{figure}

\newpage

\newpage

\begin{deluxetable}{llccc}
\tablecaption{Fitting Results of Sgr B2 to a Phenomenological Spectral Model}
\tablehead{
\colhead{Model Components}   &\colhead{Parameters}   &\colhead{Unit} 
&\colhead{GIS} &\colhead{SIS}}
\startdata
Absorption     & $N_{\rm H}$\tablenotemark{a} & (H cm$^{-2}$)
&8.3$^{+2.5}_{-2.0} \times 10^{23}$&$8.3 \times 10^{23}$ (fixed)\nl
Continuum       & Photon Index              &     &2.0
(fixed)&2.0 (fixed) \nl
                & Flux\tablenotemark{b} (4--10 keV) & (ph s$^{-1}$ cm$^{-2}$)
      &1.5 $^{+0.3}_{-0.2} \times 10^{-4}$&1.8 $^{+0.6}_{-0.6} \times 10^{-4}$
 \nl
Fe 6.4-keV Line & Flux\tablenotemark{b} & (ph s$^{-1}$ cm$^{-2}$)
&9.7 $^{+1.0}_{-1.0} \times 10^{-5}$&9.7 $^{+1.7}_{-1.7} \times 10^{-5}$ \nl
& Equivalent Width & (keV) &2.9$^{+0.3}_{-0.9}$&2.1 $^{+2.1}_{-0.8}$ \nl
\tableline
Total Luminosity& $L_{4-10 {\rm keV}}$ & (erg s$^{-1}$)
& $1.1 ^{+0.7}_{-0.5} \times 10^{35}$&$1.4 ^{+0.7}_{-0.9} \times 10^{35}$\nl 
\tableline
Reduced $\chi^2$ (d.o.f.)& &
&0.91 (34)    &0.78 (45) \nl
\enddata
\tablecomments{The errors are at 90\% confidence level.}
\tablenotetext{a}{The equivalent hydrogen column density for the solar 
abundances.}
\tablenotetext{b}{The fluxes are not corrected for absorption.}
\end{deluxetable}

\begin{deluxetable}{lcrr} 
\tablecaption{Comparison between Required Luminosities and the Observed Ones}
\tablehead{
\colhead{X-ray Binary}&\colhead{$d$\tablenotemark{a}}&\colhead{$L_{\rm
obs}$\tablenotemark{b}}&\colhead{$L_{\rm
req}$\tablenotemark{c}} \\
\colhead{Sources}	
&\colhead{(pc)}&\colhead{(erg s$^{-1}$)} &\colhead{(erg s$^{-1}$)}}
\startdata
A 1742$-$294	&170	& $2 \times 10^{36}$	&$9 \times 10^{39}$ \nl
1E 1740.7$-$2942	&233	& $4 \times 10^{36}$	&$2 \times 10^{40}$ \nl
1E 1743.1$-$2843	&63	& $1 \times 10^{36}$	&$1 \times 10^{39}$ \nl
\enddata
\tablenotetext{a}{The projected distance to Sgr~B2.}
\tablenotetext{b}{The observed luminosity in the 2--10 keV band (Sidoli et al. 
1999\markcite{Sidoli99})}
\tablenotetext{c}{The required luminosity to account for the
observed X-ray fluxes of Sgr~B2}
\end{deluxetable}

\end{document}